\documentstyle[PASJadd]{PASJ95}
\draft 

\newcommand{\vol}{}
\newcommand{\no}{}
\newcommand{\lett}{}
\newcommand{\spage}{}
\newcommand{\rdate}{2001 February 23}
\newcommand{\adate}{2001 September 28}

\def\deg{$^\circ$ }

\begin{document}

\title{ASCA Observation of the Nearest Gravitational Lensing
Cluster Candidate -- A 3408}

\author{ %% begin:author env.
Haruyoshi {\sc Katayama}, Kiyoshi {\sc Hayashida}, and Kiyoshi {\sc
Hashimotodani} \\
% \thanks{Example: Present Address is xxxxxxxxxx} \\
{\it Graduated School of Osaka University, 1-1, Machikaneyama, Toyonaka,
Osaka 560-0043} \\
{\it hkatayam@ess.sci.osaka-u.ac.jp} 
} %% end:author env.

\abst{

 We observed the nearest gravitational lensing cluster candidate, A 3408
 (${\it z\/}=0.042$), with ASCA. The projected mass profile of A 3408
 was determined from the ICM temperature and the $\beta$-model
 parameters obtained with ASCA, assuming that the hot gas is spherically
 symmetric and in hydrostatic equilibrium. The projected mass within an
 arc radius, $r_{\rm arc}$, of 110 kpc is $M(r_{\rm
 arc})=1.2^{+0.8}_{-0.4}\times10^{13}~ \MO$.  This is 18 -- 45\% of the
 mass previously derived from a lensing analysis by Campusano et al. 
 (1998, AAA 069.160.189) without any X-ray information.

 The primary cause of this discrepancy is in their assumption that the
 center of the cluster potential coincides with the position of the
 brightest cluster galaxy (BCG), while we reveal the BCG position to be
 60$''$ outside of the X-ray center.  We further calculated a model for
 the source galaxy position and the gravitational potential that can
 reproduce both the X-ray data and the arc image.  In this model, the
 magnification factor of the lens for the source galaxy was evaluated to
 be ${\Delta}m=$ 0.07$\pm$0.03 mag; i.e., the source galaxy is slightly
 magnified by the lens cluster A 3408.
} 

\kword{galaxies: clusters: individual (A 3408) --- X-rays: galaxies:
clusters --- Xrays: individual (A 3408)}

\maketitle
\thispagestyle{headings}

\section{Introduction}

Clusters of galaxies are the largest gravitationally bound systems in
the universe, which set clear constraints on the formation of the structure
and composition of the universe. Furthermore, the gravitational
mass of clusters is an important observational quantity to constrain the
density parameter, $\Omega$.

There are primarily two methods to estimate the total (gravitational)
mass of clusters of galaxies. One method employs X-ray observations
under the assumption of hydrostatic equilibrium of the hot gas. The
other uses the configuration of gravitational lensing arcs seen in
optical observations.  However, the mass estimates from lensing observations
have been found to be systematically about 2-times larger than the mass
estimates from X-ray observations (Wu, Fung 1997). Several ideas
have been proposed to solve this discrepancy, including the following:
(1) uncertainty in selecting the cluster center (Allen 1998;
Hashimotodani 1999); (2) effect of the cooling flow (Allen 1998); and (3)
mass profiles different from the conventional $\beta$ model, e.g., the
NFW density profile (Navarro et al. 1996).

In order to examine these possibilities, detailed X-ray observations of
the central region of clusters are important.  Such observations may be
realized either with higher resolution instruments or equivalently by
taking low-{\it z\/} gravitational lensing clusters as targets.  This is 
our motivation to observe low-{\it z\/} gravitational lensing
clusters.

Recently, several groups have found gravitational lensing clusters with
low-{\it z\/}. For instance, Allen, Fabian, and Kneib (1996) discovered a {\it z\/} = 0.43 lensed arc in the massive cooling flow cluster
PKS 0745--191 at {\it z\/} = 0.103. Blakeslee and Metzger (1999)
discovered a lensed arc at {\it z\/} = 0.573 in the nearby cD cluster
A 2124 at {\it z\/} = 0.066. However, the lensing properties of these
low-{\it z\/} gravitational lensing clusters remain unclear. Especially,
detailed X-ray observations of low-{\it z\/} gravitational lensing
clusters have rarely been performed. A 3408 at {\it z\/} = 0.042 is the
lowest-{\it z\/} gravitational lensing cluster candidate in which
arclike objects are found. In this paper, we present the results of
an X-ray observation of A 3408 with ASCA.

For our study, we adopt $\Omega_0=1$, $\lambda=0$, and $H_0=50h^{-1}_{50}$
km s$^{-1}$ Mpc$^{-1}$; 1$'$ corresponds to 68 $h^{-1}_{50}$ kpc at the
distance of the cluster.

\section{Discovery of an Arc-like Object in A 3408}

Because A 3408 has a relatively low galactic latitude (${\sl
b}=-17.57^{\circ}$), detailed optical observations have not been
performed.  According to the ACO catalog (Abell et al. 1989), the number
of galaxies is 43 and the richness class is 0. Campusano and Hardy
(1996) discovered an arclike object near the center of A 3408.  Figure 1
displays the {\it B} + {\it I} band image of A 3408. An arc-like object
is $\sim50''$ away from the brightest cluster galaxy (BCG). The length
of the arc-like object is $\sim11''$, but its width is unresolved. The
central elliptical galaxy has a redshift of 0.042, and the arc-like
object has a redshift of 0.073. If the arc-like object is the lensed
image of a background galaxy, A 3408 is the nearest gravitational
lensing cluster that has been observed.

\section{Lens Models of A 3408 }

Campusano et al. (1998) considered two lensing mass models to
evaluate the projected mass within the arc radius, regarding the BCG as
the center of the potential of A 3408. \\
{\sl The minimum mass case (The BCG model)}: \\
The mass distribution follows the light profile of the central
brightest elliptical galaxy. The cluster component was not taken into
account. The mass enclosed within $r_{\rm arc}$
(from the BCG to the arc) is  $M(r_{\rm arc})\sim1.5\times10^{12}~\MO$, 
and the amplification of the source galaxy is ${\Delta}m=0.01$. \\
{\sl The maximum mass case (The dark-halo model)}: \\ 
A typical massive dark-halo of cluster scale is assumed. They adopted
a massive mass profile comparable to that found in A 2218. The mass
enclosed within $r_{\rm arc}$ is $M(r_{\rm arc})\sim4.4\times10^{13}~\MO$, and
the amplification of the source galaxy is ${\Delta}m=0.60$. 

Note that these models are based almost solely on a single arc image
and that no X-ray information was employed. They could not determine which
model is appropriate for the arc-like object.

\section{ASCA Observations of A 3408}

We observed A 3408 on 1999 March 13--14 with ASCA (Tanaka et
al. 1994). With the exception of the ROSAT all-sky survey, no other
X-ray observations had been carried out for this cluster. The data from
the Gas Imaging Spectrometer (GIS; Ohashi et al. 1996; Makishima et
al. 1996) were taken in the PH mode, while those of the Solid-state
Imaging Spectrometer (SIS) were acquired in the 2 CCD Faint mode. Because
the SIS spectrum was severely distorted, possibly by a dark current, and
its field of view is smaller than that of the GIS, we concentrated on
the GIS data in a following analysis.  The data selection was done
using the standard screening parameters. We excluded the GIS data with a
cutoff rigidity lower than 6 GV and those with elevation angles below
5\deg from the night-earth rim and 25\deg from the day-earth rim. After
screening, the total exposure time was $\simeq$ 50 ks for the GIS.

Figure 2 shows the 0.7--10 keV GIS image of A 3408 (contour) overlaid on
an optical image from the Digitalized Sky Survey (DSS).  Extended
emission from A 3408 is surely detected. The sky coordinates of the GIS
image are adjusted to the point source, Seyfert 1 galaxy 1H0707--495,
located at $15'$ south to the center of the GIS field of view. It is
found that the X-ray surface brightness peak of A 3408 is not coincident
with the position of the BCG. The X-ray peak is $\sim60''$ away from the
BCG (PGC 070830.4--491246). Considering a GIS pointing accuracy of
$24''$ (Gotthelf et al. 2000), we conclude that the BCG is not
located at the X-ray peak, which we usually assume to be the center of
the potential.

\section{Spectral Analysis}

We estimated the background spectrum by the following method. The
non-X-ray background was reproduced with a method introduced by Ishisaki
(1996). The cosmic X-ray background was extracted from the Large Sky
Survey (Ueda et al. 1999) observed during the ASCA PV phase.

We fitted the GIS spectra with a single-temperature thermal-emission model
by Raymond and Smith (1977). We fixed the hydrogen column density to the
galactic absorption of $N_{{\rm H}}=6.0\times10^{20}$ cm$^{-2}$ (Dickey, 
Lockman 1990) and the redshift to $z=0.042$. We first extracted
spectra from the region $r<250$ kpc and $250$ kpc $<r<800$ kpc. The fit
results are summarized in table 1.

The temperature of the region $r<250$ kpc is similar to that of the
region $250$ kpc $<r<800$ kpc. We also fitted the spectrum of the
region within the arc radius ($\sim$110 kpc). The temperature of the
region $r<r_{\rm arc}$ is $3.0^{+0.3}_{-0.3}$ keV. These results imply a
nearly isothermal nature for the ICM of A 3408.

We next extracted spectra from the combined region $r<800$ kpc. Figure
3 shows the spectrum of the region $r<800$ kpc. The temperature of
this region is $2.9^{+0.2}_{-0.2}$ keV and the luminosity in the 2--10
keV energy band is $L_{\rm X} = 2.3^{+0.1}_{-0.2}\times10^{43}$ $h^{-2}_{50}$
erg s$^{-1}$. The luminosity measured by the ROSAT all-sky survey
in the 0.1--2.4 keV energy band is $L_{\rm X} = 5\times10^{43}$ $h^{-2}_{50}$
erg s$^{-1}$ (Ebeling et al. 1996), which is consistent with our
result derived from the spectral fitting ($L_{\rm X} = 4.5^{+0.3}_{-0.1}\times10^{43}$ $h^{-2}_{50}$ erg s$^{-1}$). The
temperature of the ICM corresponds to a velocity dispersion of
$674^{+22}_{-24}$ km s$^{-1}$, whereas we assume equal specific energy in
the gas and galaxies.

\section{Spatial Analysis}
  
 The X-ray surface brightness profile was extracted by accumulating
 photons within annular bins of $0'.75$ width. 
 We considered the X-ray peak as the
 center.  Figure 4 shows the X-ray surface brightness profile of A 3408.
 We fitted the X-ray surface brightness profile with an isothermal
 $\beta$ model.  In the $\beta$ model, the gas density profile is described as
 $\rho_{\rm gas}(r)=\rho_{\rm gas0}/[1+(r/r_{\rm c})^2]^{\frac{3\beta}{2}}$, where
 $\beta$ is the slope parameter of the outer part of the distribution and
 $r_{\rm c}$ is the core radius of the cluster. Assuming this density
 profile, the X-ray surface brightness profile is described as

 % equation 1
 \begin{equation}
 S(b) = S_{0}\frac{1}{(1+(\frac{b}{r_{\rm c}})^{2})^{3\beta-1/2}} + {\rm background},
 \end{equation}

 \noindent where $S_{0}$ is the central surface brightness and $b$ is the
 projected 2D distance from the cluster center. Because of the
 complicated PSF of the ASCA XRT, we used a ray-tracing simulation
 (Ikebe et al. 1996) to fit the radial profile.  The fit results
 are shown in figure 4 and table 2. The profile is well fitted with a
 single $\beta$ model and doesn't exhibit a significant excess X-ray
 emission from the central region, which is expected for a cooling
 flow. Note that the core radius and $\beta$ are similar to the typical
 value of nearby clusters determined by Einstein observations
 (Abramopoulos, Ku 1983).

 We also fitted the same profile with an isothermal gas model in the NFW
 potential. The gas-density profile is analytically modeled as
 $\rho_{\rm gas}(r)=\rho_{\rm gas0}/[1+(r/r_{\rm s})]^{B\frac{r_{\rm s}}{r}}$ (Makino et
 al. 1998). Here, $r_{\rm s}$ is the scaling parameter of the cluster, and
 $B$ is the gradient of the gas density. The fit results are shown in
 figure 4 and table 3. There is little difference between two best-fit
 curves, and the $\chi^2$ value of the NFW fit is similar to that of the 
 $\beta$ fit.  
 Instruments with a better spatial resolution and a larger effective area
 are needed to distinguish these models. In this paper, we adopt the
 $\beta$-model parameters to derive the mass distribution.

\section{Mass Profile of A 3408}

 From the $\beta$-model parameters and the temperature of the ICM, we
 calculated the mass distribution of A 3408. Assuming hydrostatic
 equilibrium and neglecting the temperature gradient, the mass contained
 within a circle of radius $b$ on the sky plane is given by

 % equation 2
\begin{equation}
 M(b) = 1.14 \times 10^{14} \beta \tilde{m}(b)(\frac{kT}{\rm keV})(\frac{r_{\rm c}}{\rm Mpc})~\MO,
\end{equation}
where
 % equation 3
\begin{equation}
 \tilde{m}(b) =  \frac{R_0^{3}}{1+R_0^{2}}-\int_{b_0}^{R_0}x\sqrt{x^2-b_0^2}\frac{3+x^2}{(1+x^2)^2}dx.
\end{equation}

In equation (3), $b_0=b/r_c$ and $R_0=R/r_{\rm c}$, where $R$ is the
physical size of the cluster, for which we assumed 5 Mpc.

In order to estimate the uncertainty in our mass calculation, we
employed the 90\% confidence errors in the spectral and imaging
parameters. Figure 5 shows the mass profile of A 3408 integrated along
the line of sight. Each of the 27 lines corresponds to a combination of
either the best-fit value or the 90\% confidence lower or upper limit
of the three parameters: $\beta$, $r_{\rm c}$ and $kT$. The
projected mass within the arc radius, $r_{\rm arc}$, of 110 kpc is
$M(r_{\rm arc})=1.2^{+0.8}_{-0.4}\times10^{13}~\MO$.

We also applied polytropic models (Sarazin 1988) in which some
temperature gradient in the intracluster gas is assumed. If the hot
gas is strictly adiabatic, its pressure and density have a simple
relation of $P\propto\rho^{\gamma}$. In this model,
$\gamma=1$ implies that the gas distribution is isothermal. The
hydrostatic mass contained within a sphere of a radius $r$ is
prescribed by

% equation4
\begin{equation}
 M(r)=-\frac{\gamma kT_{\rm gas}(r)}{G{\mu}m_{\rm p}}r\frac{d{\rm ln}\rho_{\rm gas}}{d{\rm ln}r},
\end{equation}
where
% equation5
\begin{equation}
 T_{\rm gas}(r)=T_{0,\gamma}(\frac{\rho_{\rm gas}(r)}{\rho_0})^{\gamma-1}=T_{0,\gamma}(1+\frac{r^2}{r_{\rm c}^2})^{-3\beta(\gamma-1)/2}.
\end{equation}

Markevitch et al. (1998) found universal temperature gradients in nearby
rich clusters observed with ASCA, which are well described by the
polytropic distribution with $\gamma=$1.2 -- 1.3, implying temperature
increases toward the center. Therefore, we introduced a polytropic gas
distribution with $\gamma=1.3$.  Although there is no observational
evidence, we also tested the case with $\gamma=0.9$ for a temperature
decrease toward the center. However, because the temperature gradient is
significant in the region $r>r_{\rm c}$, the range of the estimated mass
is almost the same as that in the isothermal case, $M(r_{\rm
arc})=1.2^{+0.9}_{-0.4}{\times}10^{13}~\MO$ ($\gamma=0.9$) and $M(r_{\rm
arc})=1.4^{+0.6}_{-0.5}{\times}10^{13}~\MO$ ($\gamma=1.3$).

\section{Discussion}

 \subsection{Comparison with Models by Campusano et al.}

 The existence of hot gas extending over a few Mpc in A 3408 rejects the
 BCG model by Campusano et al. (1998), in which only the BCG was taken
 into account.  The projected mass of A 3408 within the arc radius
 derived based on our X-ray observation is
 $1.2^{+0.8}_{-0.4}\times10^{13}~\MO$.  This is 18 -- 45\% of the mass
 obtained from the dark-halo (cluster) model by Campusano et al. (1998). 
 However, it should not be taken straightforward, since the center of
 the potential that they assumed, i.e., BCG, differs from the X-ray
 center that we defined.  Furthermore, they assumed that the cluster
 mass profile of A 3408 is similar to that of a distant cluster, A 2218,
 while our mass profile is based on the X-ray observation of A 3408,
 itself. On these two points, we consider that our mass estimation based 
 on the X-ray observation has some advantage over the mass estimation by
 Campusano et al. (1998).

 Our mass model is based on the following assumptions: an isothermal gas
 distribution, hydrostatic equilibrium, and a spherically symmetric gas
 distribution. However, as mentioned in section 7, the range of
 the estimated mass is little affected by possible temperature gradients
 with $\gamma$=1.3 or 0.9. Meanwhile, if A 3408 is a merging or
 post-merging cluster, the assumptions of hydrostatic equilibrium and
 a spherically symmetric gas distribution would be inappropriate. However,
 the relatively regular X-ray surface-brightness distribution of A 3408
 suggests that the cluster is a moderately relaxed system. This is also
 supported by the fact that the morphological classification of A 3408 is
 Bautz-Morgan Type I-II (Abell et al. 1989). In order to
 confirm that A 3408 is really a relaxed system, a further investigation, such
 as concerning the velocity dispersion of member galaxies, is needed.

 \subsection{Potential Model Accounting for Both the X-ray Data and the 
 Arc Image}

 For a given gravitational potential (cluster mass distribution) and a
 source position, the X-ray emission profile and the lensing arc image
 can be calculated.  Conversely, we constrained the source position and
 the gravitational potential that reproduce the arc image and the X-ray
 data, as was done in Hattori et al. (1998) and in Hashimotodani (1999).
 We performed a similar analysis for A 3408. In addition to the
 potential of the cluster (we allow for deviations from spherical
 symmetry), the potential due to the BCG was considered, since, in
 several clusters of galaxies, bright galaxies near the arc are
 considered to play a significant role in lensing (Hattori et al. 1998,
 Hashimotodani 1999). The potential depth of the BCG was estimated by
 using the Faber--Jackson relation (Faber, Jackson 1976); it is
 $\sim300$ km s$^{-1}$ in terms of the velocity dispersion. Figure 6
 shows the source position predicted by our model. The source position
 determined by Campusano et al. (1998) in their dark halo model is shown
 in the same figure as a reference. The magnification factor for the
 source galaxy is constrained to be ${\Delta}m=0.07\pm0.03$ mag. Thus,
 the source galaxy is only slightly magnified by the lens A 3408 in our
 potential model. The small magnification factor is also consistent with
 the slightly distorted image of the arc-like object.

 Because of the small redshift of the source galaxy for the A 3408 lens,
 we have a good opportunity to study its intrinsic properties and the
 modification by the lens in detail. The characteristics of the source
 galaxy were discussed by Campusano et al. (1998) under the assumptions
 of their minimum mass case and the maximum mass case, for which they
 evaluated the source magnification factors to be ${\Delta}m=0.01$ and
 0.60, respectively. On the other hand, we provided ${\Delta}m=0.07$,
 which implies that the absolute magnitude, $M_B$, of the source galaxy
 is $-18.7$.  According to the galaxy population statistics by Binggeli,
 Sandage, and Tammann (1988), about 75\% of the field galaxies with the
 same absolute magnitude are spiral. Considering the small magnification
 factor, ${\Delta}m=0.07$, of the lens, the source galaxy is not
 significantly modified in its shape, and should be regarded as an
 edge-on spiral originally. In this case, except for the small
 probability to realize an edge-on configuration, the low rotation
 velocity of the source galaxy measured by Campusano et al. (1998) might
 imply a somewhat extraordinary nature of the source galaxy. This is
 because even the smallest Sc galaxies have twice the observed rotation
 velocity of the source galaxy, as mentioned by Campusano et al. (1998). 
 Although further study is required concerning this point, this kind of
 examination is generally important for a nearby lensing object.

\section{Summary}

 We observed the nearest gravitational lensing cluster candidate, A 3408,
 with ASCA.  Our observation determined the ICM temperature and
 X-ray surface brightness profile of A 3408 for the first time. We
 also showed that the gas distribution of A 3408 is isothermal at the
 central region of A 3408, even within the arc radius.

 We compared our mass model based on the X-ray observation with that of
 Campusano et al. (1998) based on the optical image. The mass derived by
 us is 18 -- 45\% of that by Campusano et al. (1998). The primary cause
 of this discrepancy is in their assumption that the center of the
 cluster potential coincides with the position of the BCG, while we
 reveal the BCG position to be 60$''$ outside of the X-ray center.

 As shown in this paper, X-ray observations are essential to study
 gravitational lens phenomena and to determine mass profiles in clusters
 of galaxies.  Hashimotodani (1999) found that there is a significant
 number of samples in which the BCG position deviates from the X-ray
 peak of the cluster. He suggested that this can be one of the primary
 causes for the discrepancy in the mass estimates from the X-ray data and
 from the lensing images.

 One advantage of observing low-{\it z} gravitational lensing
 clusters is that it enables us to study X-ray profiles even with an
 instrument having only moderate spatial resolution. Further detailed studies
 of the central part of A 3408, such as a temperature gradient and gas
 distribution within the arc radius, will be enabled with better spatial
 resolution and larger effective-area instruments, like XMM-Newton.
 Another advantage is that it becomes possible to examine a lensed image of
 the source galaxy, if the source redshift is small, as in our case. In
 fact, we showed that the weak shear of the arc-like object is
 consistent with the small magnification factor which we derived.

 We are grateful to the ASCA team for operating the spacecraft and
 supporting the data analysis.  H.K. is supported by JSPS Research
 Fellowship for Young Scientists.  The computation code for the lens 
 image and cluster potential used in the last subsection was 
 provided by M. Hattori and M. Murata, Tohoku University.

\section*{References}

%%%
% See the manual for the detail.
%%%
\small
 \re
 Abell, G. O., Corwin, H. G., Jr. \& Olowin, R. P.\ 1989, ApJS,\ 70, 1 
 \re
 Abramopoulos, F., \& Ku, W. H. -M.\ 1983, ApJ,\ 271, 446
 \re
 Allen, S. W.\ 1998, MNRAS,\ 296, 392
 \re
 Allen, S. W., Fabian, A. C., \& Kneib, J. P.\ 1996, MNRAS,\ 279, 615
 \re
 Binggeli, B., Sandage, A., \& Tammann, G. A.\ 1988, ARA\&A,\ 26, 509
 \re
 Blakeslee, J. P., \& Metzger, M. R.\ 1999, ApJ,\ 513, 592
 \re
 Campusano, L. E., \& Hardy, E.\ 1996, in IAU Symp,\ 173, Astrophysical
 applications of gravitational lensing, ed. C. S. Kochanek \& J. N. Hewitt (Dordrecht: Kluwer), 125
 \re
 Campusano, L. E., Kneib, J. -P., \& Hardy, E.\ 1998, ApJ,\ 496, L79
 \re
 Dickey, J. M., \& Lockman, F. J.\ 1990, ARA\&A,\ 28, 215 
 \re
 Ebeling, H., Voges, W., B$\ddot{o}$hringer, H., Edge, A. C., Huchra,
 J. P., \& Briel, U. G.\ 1996, MNRAS, \ 281, 799
 \re
 Faber, S. M., \& Jackson, R. E.\ 1976, ApJ,\ 204, 668 
 \re
 Gotthelf, E. V., Ueda, Y., Fujimoto, R., Kii, T., \& Yamaoka, K.\ 2000, ApJ,\ 543, 417
 \re
 Hashimotodani, K.\ 1999, Ph. D. Thesis, The University of Osaka
 \re
 Hattori, M., Matuzawa, H., Morikawa, K., Kneib, J. -P.,
 Yamashita, K., Watanabe, K., B$\ddot{o}$hringer, H., \& Tsuru, G. T.\
 1998, ApJ,\ 503, 593  
 \re
 Ikebe, Y.\ 1996, Ph. D. Thesis, The University of Tokyo
 (RIKEN IPCR CR-87)
 \re
 Ishisaki, Y.\ 1996, Ph. D. Thesis, The University of Tokyo (ISAS-RN 613)
 \re
 Makino, N., Sasaki, S., \& Suto, Y.\ 1998, ApJ,\ 497, 555
 \re
 Makishima, K., Tashiro, M., Ebisawa, K., Ezawa, H., Fukazawa, Y., Gunji, S.,
 Hirayama, M., Idesawa, E. et al.\ 1996, PASJ,\ 48, 171
 \re
 Markevitch, M., Forman, W. R., Sarazin, C. L., \& Vikhlinin, A.\ 1998, 
 ApJ,\ 503, 77
 \re
 Navarro, J. F., Frenk, C. S., \& White, S. D. M.\ 1996, ApJ,\ 462, 563
 \re
 Ohashi, T., Ebisawa, K., Fukazawa, Y., Hiyoshi, K., Horii, M., Ikebe, Y.,
 Ikeda, H., Inoue, H. et al.\ 1996, PASJ,\ 48, 157
 \re
 Raymond, J. C., \& Smith, B. W.\ 1977, ApJS,\ 35, 419
 \re
 Sarazin, C. L.\ 1988,``X-ray emissions from clusters of galaxies'',
 (Cambridge:  Cambridge University press)
 \re
 Tanaka, Y., Inoue, H., \& Holt, S. S.\ 1994, PASJ,\ 46, L37
 \re
 Ueda, Y., Takahashi, T., Inoue, H., Tsuru, T., Sakano, M., Ishisaki, Y.,
 Ogasaka, Y., Makishima, K. et al.\ 1999, ApJ,\ 518, 656
 \re
 Wu, X. -P., \& Fung, L. -Z.\ 1997, ApJ,\ 483, 62 
\\

\label{last}

%\newpage
\clearpage

%%%%%%%%%%%
% Table 1

\begin{table*}
\begin{center}
Table~1.\hspace{4pt}Spectral fits to the GIS spectra of three regions of A 3408.\\
\end{center}
\vspace{6pt}
 \begin{tabular*}{\textwidth}{l c c c } \hline\hline\\[-6pt]
 Region (kpc) & Temperature (keV) & Abundance (solar) & $\chi^2$/d.o.f \\ [4pt]\hline\\[-6pt]
  $r<250$ &  $3.0^{+0.3}_{-0.2}$ & $0.42^{+0.15}_{-0.21}$ &
 143.0/119 \\
 $250<r<800$& $2.8^{+0.2}_{-0.3}$ & $0.28^{+0.18}_{-0.22}$ &
 252.6/229 \\ 
 $r<800$& $2.9^{+0.2}_{-0.2}$ & $0.34^{+0.20}_{-0.17}$ &
 245.5/238 \\ \hline
 \end{tabular*}
\vspace{6pt}
\end{table*}
\newpage

%%%%%%%%%%%
% Table 2
\begin{table*}
\small
\begin{center}
Table~2.\hspace{4pt}Radial surface-brightness fit to the GIS
 data with the $\beta$ model.\\
\end{center}
\vspace{6pt}
   \begin{tabular*}{\textwidth}{c c c c c}\hline\hline\\[-6pt]
$\beta$ & $r_{\rm c}$ ($h_{50}^{-1}$ kpc) & $S_0$ (count/pix)  & background (count/pix)   & $\chi^2$/d.o.f  \\\hline
 $0.53\pm0.07 $ & $177\pm61$ &
    6.55$\pm$0.08  & 0.45$\pm$0.01 & 41.6/19 \\\hline
% &  & & &  \\
   \end{tabular*}
\vspace{6pt}
\end{table*}

%%%%%%%%%%%
% Table 3
\begin{table*}
\small
\begin{center}
Table~3.\hspace{4pt}Radial surface-brightness fit to the GIS
 data with the NFW model fitting.\\
\end{center}
\vspace{6pt}
   \begin{tabular*}{\textwidth}{c c c}\hline\hline\\[-6pt]
    $B$ & $r_{\rm s}$  ($h_{50}^{-1}$ kpc) & $\chi^2$/d.o.f \\\hline 
    $7.7\pm1.0$ & $632\pm204$  & 43.6/19\\ \hline
   \end{tabular*}
\vspace{6pt}
\end{table*}

%\newpage
\clearpage

\centerline{Figure Captions}

% Fig.1
\begin{fv}{1}
 {0cm}
 {{\it B} + {\it I} band image of an arc-like object in A 3408 taken with the
0.9 m CTIO telescope (Campusano et al. 1998). The image size is
 $2'\times2'$. An arc-like object is seen 50$''$ away from the BCG.}
\end{fv}

% Fig.2
\begin{fv}{2}
{0cm}
 {GIS image of A 3408 (contour) overlaid on
  an optical image from DSS (Digitalized Sky Survey). The GIS energy
 range is 0.7--10 keV. The GIS image was
  smoothed by a $\sigma=0^{'}_{.}5$ Gaussian  filter. The spacing of the 
  contour is linear with a step size of $4.2\times10^{-4}$
  count s$^{-1}$ arcmin$^{-2}$. The lowest contour is at $8.6\times10^{-4}$
  count s$^{-1}$ arcmin$^{-2}$. The circle represents the GIS pointing accuracy.}
\end{fv}

% Fig.3
\begin{fv}{3}
 {0cm}
 {GIS spectrum of the region $r<800$ kpc.}
\end{fv}

% Fig.4
\begin{fv}{4}
 {0cm} 
{X-ray surface brightness profile of A 3408. The solid line is the
 best-fit curve of the $\beta$-model fitting and the dotted line is that
 of the NFW model fitting.}
\end{fv}

% Fig.5
 \begin{fv}{5}
  {0cm}
{Projected mass distribution of A 3408. }
 \end{fv}

% Fig.6
 \begin{fv}{6}
  {0cm} {Magnifications of a background source at ${\it z\/}=0.073$
  produced by our lens model of A 3408. The image size is
  $100''\times100''$, and the center (0,0) corresponds to the BCG
  position. The contours indicate magnifications of ${\Delta}m=$0.04, 0.05,
  0.07, 0.11, and 0.19 from the outer to the inner. S1 and S2 correspond
  to the position of the source galaxy derived by Campusano et
  al. (1998) and by our model, respectively. The cross represents the
  X-ray peak.}
  \end{fv}

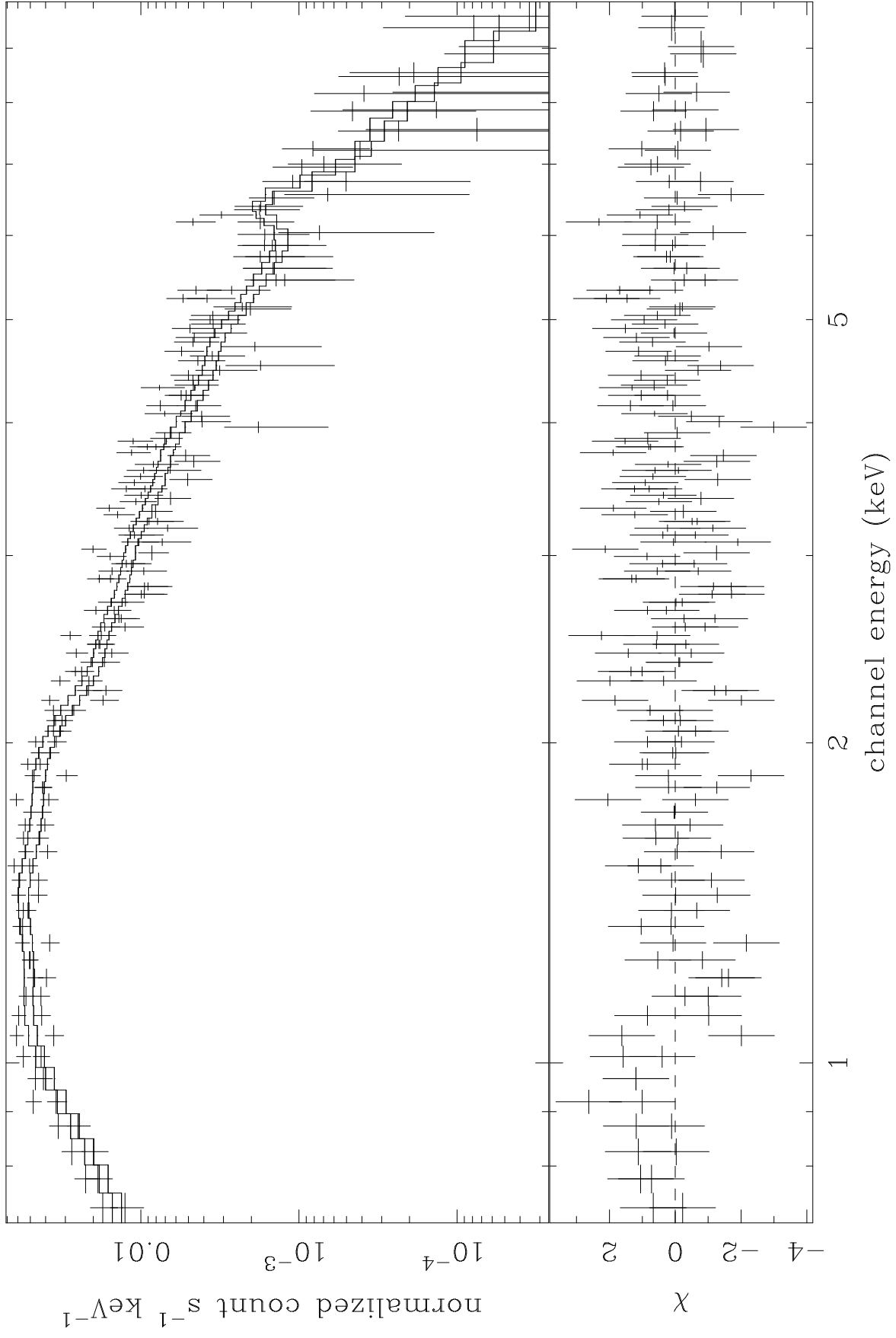
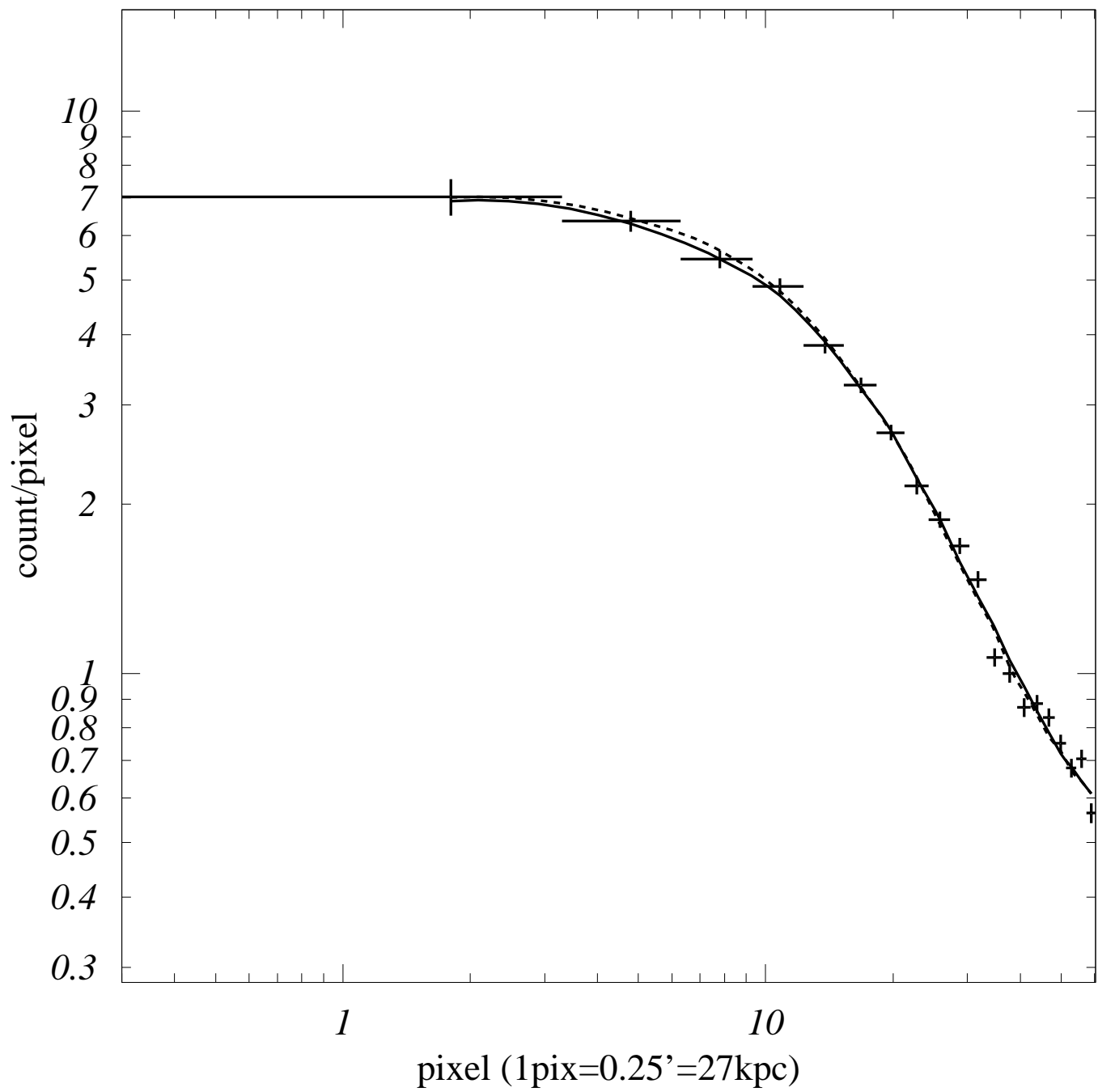
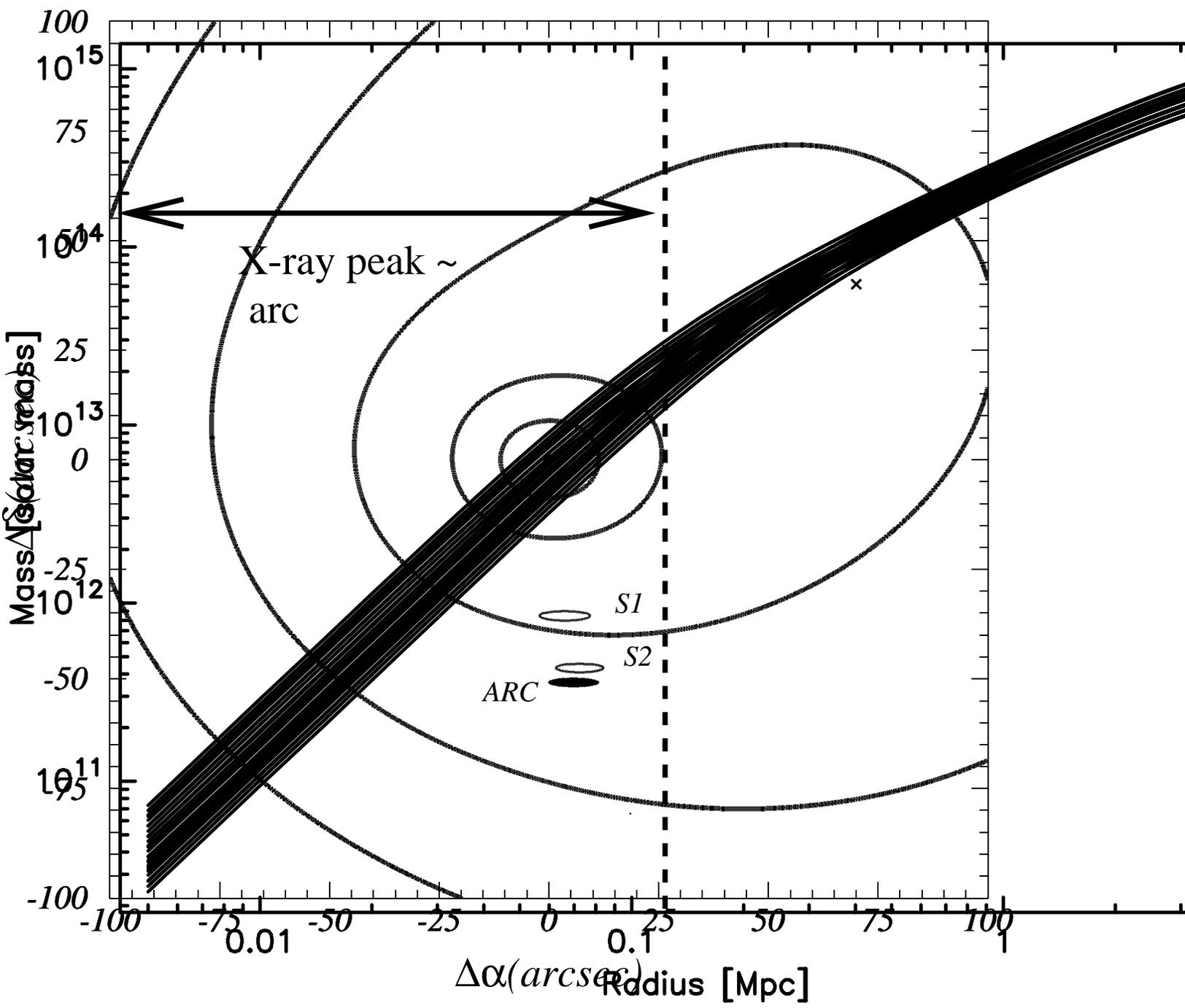
\end{document}